%% file: main.tex
\documentclass[twocolumn]{aastex701}
\graphicspath{{figures/}{./}}
\usepackage{amsmath}
\usepackage{mathrsfs}
\usepackage{soul}

\newcommand\gt{>}

\newcommand{\kms}{\ {\rm km\ s}^{-1}}

\definecolor{codegreen}{rgb}{0,0.6,0}
\definecolor{codegray}{rgb}{0.5,0.5,0.5}
\definecolor{codepurple}{rgb}{0.58,0,0.82}
\definecolor{backcolour}{rgb}{0.95,0.95,0.92}

\definecolor{mlp}{rgb}{0.294, 0.612, 0.827}

\newcommand{\FI}{Center for Computational Astrophysics, Flatiron Institute, 162 Fifth Avenue, New York, NY, USA\\ }

\shorttitle{Broadening Kernels and Formation Temperatures}
\shortauthors{Palumbo}

\begin{document}

\title{The Limits of Line Broadening: \\ Modeling Stellar Spectra and Formation Temperatures at High Resolution}


\correspondingauthor{Michael L. Palumbo III}
\email{mpalumbo@flatironinstitute.org}

\author[0000-0002-4677-8796]{Michael L. Palumbo III}
\email{mpalumbo@flatironinstitute.org}
\affiliation{\FI}

\begin{abstract}
The modeling of stellar spectra is pervasive in astronomy. Conventionally, the shapes of absorption lines are modeled by convolving thermal profiles (computed given some model stellar atmosphere and line list) with broadening kernels intended to account for the effects of rotation and other nonthermal sources of broadening (i.e., macroturbulence). Here, we show that the assumptions that permit this convolution can break down at high spectral resolution and produce appreciable errors in the modeled flux. We then consider the effects of rotation, microturbulence, and macroturbulence on the intensity and flux contribution functions, which astronomers use to map individual spectral segments to quasi-physical formation ``locations'' in the stellar atmosphere. We show that proper consideration of (1) the distinction between intensity and flux and (2) the inclusion of rotation and macroturbulence in the contribution function can dramatically change the modeled formation temperatures. To complement this analysis, we provide a package --- \texttt{FormationTemperatures.jl} --- which quickly computes model line contribution functions and formation parameters given bulk stellar properties as input. In closing, we emphasize the assumptions inherent to this analysis, consider in which regimes the convolution expression for flux should be avoided, and caution how the concept of a singular ``formation temperature'' can oversimplify some realities of radiative transfer. 
\end{abstract}


\section{Introduction} \label{sec:intro}

Science cases across many distinct subdomains of astronomy necessitate the detailed modeling of stellar spectra, including stellar population synthesis modeling in galactic evolution codes \citep[e.g.,][]{Leitherer1999}, abundance measurements for galactic archeology \citep[e.g.,][]{GarciaPerez2016, Behmard2025}, and more recently in radial velocity (RV) exoplanet science for characterizing intrinsic stellar variability \citep[e.g.,][]{AlMoulla2022, AlMoulla2024, AnnaJohn2025}. In many of these applications, relatively simple model spectra are generated given some model stellar atmosphere and line list, often (but not always) under the assumption of local thermodynamic equilibrium (LTE). These models are generally inexpensive to compute, but consequently often fail to precisely model line shapes, which are broadened and skewed by convective processes in stellar atmospheres \citep[][]{Gray1982}. As a further complication, line lists are generally incomplete and atomic parameters are not perfectly known \citep[][]{Johansson2003}. \par 

Various solutions have been devised to cope with these issues. For example, it is common practice in differential analyses of abundances to manually tune line oscillator strengths to produce line profiles that better match those in an observed stellar spectrum. Since these tuned parameters are then held constant across the stellar sample in question, any systematic errors are effectively calibrated out \citep[see, e.g.,][]{Reggiani2016}. \par 

In comparison to chemical analyses, RV science is somewhat insulated from these pitfalls in stellar spectral modeling. This is because, generally speaking, RV measurements themselves do not rely on model spectra, and are instead measured from the apparent Doppler shift of the observed spectrum relative to some reference spectrum or template ``mask'' \citep[see][]{Bouchy2001, Pepe2002}. However, various stellar physical processes (including convection, pressure modes, and magnetic activity) perturb the positions and shapes of spectral lines used to measure velocities, masking the true center-of-mass movement of the star (i.e., the motion incurred by planetary companions; \citealt{Crass2021}). In an attempt to better characterize this stellar variability, recent investigations have begun using model spectra in their analyses. \par 

In one such work, \citet{AlMoulla2022} devised a method for modeling the ``formation temperature'' (which they dub $T_{1/2}$) of individual wavelength elements in a stellar spectrum. Using this method, they showed that the RV variability of spectral regions itself varies over different temperatures (or equivalently, heights) in the stellar photosphere. Subsequent works have adopted this methodology to constrain the impact of stellar granulation on different absorption lines or regions of the spectrum. For example, \citet{AnnaJohn2025}, modeled formation temperatures for an especially quiet K dwarf (HD 166620; see \citealt{Baum2022} and \citealt{Luhn2022}) and again found differences in the RV variability between hotter and cooler temperature bins, which they attribute to granulation. \par 

In these works, the authors have taken appropriate care not to interpret their modeled formation temperatures in an absolute sense. Rather, they compare the temporal RV variability of spectral regions in different modeled temperature bins, using only spectral regions where their synthesis code reproduced the observed line shapes in a reference stellar spectrum. This approach avoids one major pitfall inherent to the nature of the formation temperature approach: $T_{1/2}$ is \textit{not} directly measured from an observed spectrum, but rather it is a theoretical quantity that is dependent on the model stellar atmosphere and line list(s) used. \par 

Despite this diligence in the treatment of $T_{1/2}$, the formation temperature methodology as documented in the literature to date contains inconsistencies with respect to the definitions of intensity and flux. 
Moreover, these calculations rely on assumptions (such as achromatic limb darkening and solid-body rotation) that can meaningfully affect the shape of the model spectrum at the spectral resolving powers achieved by extremely precise radial velocity (EPRV) spectrographs. 

In fact, the issues present in the formation temperature literature warrant reconsideration of certain modeling assumptions made in synthesizing stellar spectra, especially at extremely high resolution and precision. Modern spectrographs boast resolving powers $R$ of $10^5$ and greater, in which regime the intrinsic stellar broadening (from rotation, turbulence, etc.) that sculpts line shapes is no longer dwarfed by the instrumental broadening. In this work, we aim to examine these assumptions and quantify the deviations that they produce in the modeled flux and formation temperature spectra. \par 


In \S\ref{radiative_transfer}, we review the relevant physics of radiative transfer and show how approximations made in modeling additional nonthermal broadening \added{(from rotation, microturbulence, and macroturbulence)} can fail in flux computations. In \S\ref{contribution_functions}, we consider the effects of broadening on line contribution functions, which then propagate to the modeled formation temperatures. In \S\ref{discussion}, we discuss the implications of these results for a few science cases, and consider limitations of the formation temperature model. Finally, we summarize our main points in \S\ref{conclusions}. \par  

To accompany this text and to aid astronomers in producing model formation-temperature spectra that properly account for the effects presented herein, we have written a package (\texttt{FormationTemperatures.jl}\footnote[1] {\url{https://github.com/palumbom/FormationTemps.jl}}; \citealt{FormationTemps_jl}) that wraps the spectral synthesis code \texttt{Korg.jl} \citep[][]{Wheeler2023, Wheeler2024}. \added{Instructions for calling the code from Python are included.} In addition to the source code, we also provide the scripts used to produce the figures and other numerical results presented in this work. \par 





\section{Modeling Stellar Spectra} \label{radiative_transfer}

To model a stellar spectrum, we must solve for the radiative transport of photons through the stellar atmosphere. For cool dwarf stars, we generally assume a plane-parallel, inhomogeneous atmosphere. In this case, the radiation transport equation is given by

\begin{equation} \label{eq:transport}
    \mu \frac{dI_\nu}{d \tau_\nu} =  I_\nu - S_\nu,
\end{equation}

\noindent where $\mu$ is the cosine of the angle $\theta$ subtended by the line-of-sight and surface-normal vectors,
$I_\nu$ is the monochromatic intensity at frequency $\nu$ (i.e.,\ the specific intensity\footnote{Hereafter, for the sake of brevity, we simply state ``intensity,'' ``flux,'' etc., rather than qualifying these quantities as ``specific,'' as we do not consider quantities integrated over wavelength/frequency in this work.}), $\tau_\nu$ the optical depth, and $S_\nu$ the source function \citep{Schwarzschild1906, Chandrasekhar1950}. Throughout this work, we assume LTE, in which case the source function $S_\nu$ is given by the Planck function $B_\nu$. The limitations and consequences of the above assumptions are discussed in \S\ref{sec:assumptions}. \par 

The intensity at the top of the stellar atmosphere (the ``emergent intensity,'' $I_\nu^+$) is obtained by solving Equation~\ref{eq:transport} for a ray directed outward toward the observer (i.e., $1 \geq \mu \gt 0$):

\begin{equation} \label{eq:intensity}
    I_\nu^{+}\left(\mu\right) \equiv I_\nu (\tau_\nu=0, \mu) =\int_0^{\infty} \frac{1}{\mu} S_\nu\left(t_\nu\right) \mathrm{e}^{-t_\nu / \mu} \mathrm{d} t_\nu
\end{equation}

\noindent \added{where $t_\nu$ is the $\tau_\nu$-like variable of integration.} The optical thickness, $\tau_\nu$, is expressed in terms of the total absorption coefficient\footnote{We here use the total absorption coefficient $\alpha_\nu$, rather than the opacity $\kappa_\nu$ (as in, e.g., SME; \citealt{Valenti1996}). The absorption $\alpha_\nu$ is expressed in units of per length, such that $t_\nu$ and $\tau_\nu$ are dimensionless.} $\alpha_\nu$ via

\begin{equation} \label{eq:tau_alpha}
    d \tau_\nu(s) = -\alpha_\nu(s) ds, 
\end{equation}

\noindent where $s$ is the path along the ray. 
\added{The optical depth, $\tau_\nu$,} can also be thought of as a ``coordinate'' in the stellar atmosphere, much like geometrical height (albeit one that varies with $\nu$). Typically, the specific \added{optical} thickness $\tau_\nu$ \added{at some reference wavelength (usually 5000 \AA) is provided as a coordinate in model atmospheres \citep[see also Appendix~A of][]{Wheeler2023}.} \par 

Equation~\ref{eq:intensity} describes the area-normalized specific luminous power of a ray emergent along the direction $\mu = \cos \theta$. For stars other than the Sun, we do not observe discrete portions of the surface, but rather light from the whole visible hemisphere. This quantity, the flux $\mathcal{F}_\nu$, is the integration of the line-of-sight-directed intensities over the visible hemisphere of the star. Generally, we are interested in the flux emergent from the top of the atmosphere, given by


\begin{align} \label{eq:flux}
    \mathcal{F}^+_\nu \equiv \mathcal{F}_\nu(\tau_\nu = 0) &= \int_{0}^{1} \mu I_\nu(\tau_\nu=0, \mu)d\mu \\
    \nonumber &= \int_{0}^{1} \mu \left[ \int_0^{\infty} \frac{1}{\mu} S_\nu\left(t_\nu\right) \mathrm{e}^{-t_\nu / \mu} \mathrm{d} t_\nu \right] d\mu.
\end{align}

\added{\citet{Mihalas1978b}} shows that the emergent flux $\mathcal{F}^+_\nu$ (assuming isotropic $S_\nu$) can be written in terms of a special function known as the \textit{second exponential integral} ($E_2$; \citealt{Abramowitz1972}):

\begin{equation} \label{eq:milne}
    \mathcal{F}^+_\nu = 2\pi \int_0^\infty S_\nu(t_\nu) E_2(t_\nu)dt_\nu
\end{equation}

\noindent In numerical radiative transfer, the above equation greatly simplifies and expedites the computation of stellar spectra, since it obviates the need to explicitly compute and integrate over the intensities at many discretely sampled lines of sight (i.e., values of $\mu$) in Equation~\ref{eq:flux}. \par

In practice, the above formulation produces model lines that are very narrow compared to those observed in real stellar spectra. \added{This is due to a couple effects. First, stars rotate, which Doppler shifts the local intensity profiles by the line-of-sight rotational velocities. Additionally, near-surface convection creates spatial inhomogeneity and velocity variations in the stellar atmosphere; these effects are typically modeled with ad hoc free parameters dubbed ``microturbulence'' and ``macroturbulence.''} These additional sources of broadening are normally handled by convolving the line absorption coefficients, intensity, and/or flux with broadening kernels, depending on the broadening model(s) used. We discuss these broadening sources in greater detail in \S\ref{subsec:additional_broadening}. \par 



\subsection{Modeling Additional Broadening} \label{subsec:additional_broadening}

\begin{figure} 
    \epsscale{1.125}
    \plotone{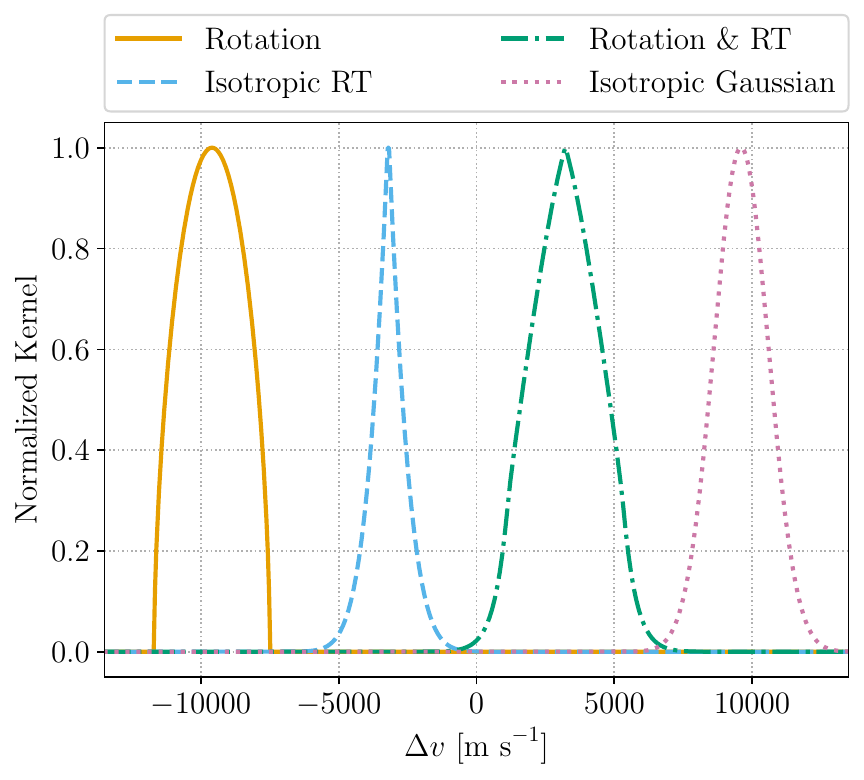}
    \caption{Example broadening kernels \added{for rotation and radial-tangential (RT) macroturbulence} from \citet{Gray2008} and \citet{Hirano2011} at arbitrary velocity offsets. The following values were adopted to generate these kernels: $v\sin i = 2.1\ \kms$, $\zeta_{\rm RT} = 1.4\ \kms$, and quadratic limb darkening coefficients $u_1 = 0.4$ and $u_2 = 0.26$. The broadening kernels have all been normalized between 0 and 1 to better illustrate their shape differences.} \label{fig:kernels}
\end{figure} 

\input{spectroscopic}

Model flux spectra include extra broadening from microturbulence, macroturbulence, and rotation in order to better reproduce the shapes and widths of lines observed in real stellar spectra. We again emphasize that micro- and macroturbulence are not exactly physically meaningful quantities, but rather additional free parameters introduced to compensate for the limitations of the assumed time-static plane-parallel model atmosphere, which is not representative of the dynamic reality of physical stellar atmospheres. \par 

These additional broadening sources can be categorized as either anisotropic or isotropic; that is, they either do or do not vary with location on the disk. Isotropic broadening is generally easier to model, as it can be analytically reduced to a simple convolution without further assumption, as we discuss in \S\ref{subsubsec:isotropic}. Anisotropic broadening, two cases of which we discuss in \S\ref{subsubsec:rot_broadening} (rotation) and \S\ref{subsubsec:rt_broad} (radial-tangential macroturbulence), does not generally permit analytical simplifications without further approximations. Here, we consider these broadening models and quantify the error introduced by relying on analytical approximations for the broadened flux spectrum. We show examples of commonly used broadening kernels (discussed further below) in Figure~\ref{fig:kernels}.  \par 

\subsubsection{Microturbulence and Isotropic Macroturbulence} \label{subsubsec:isotropic}

Microturbulence, which is understood as a turbulent velocity distribution on length scales smaller than the photon mean free path, broadens the line absorption coefficient (or alternatively, the opacity):

\begin{equation}
    \alpha^{\rm mic}_{\nu} = \alpha_\nu \circledast N(\Delta \nu),
\end{equation}

\noindent where $\circledast$ denotes the convolution operation and $N(\Delta \nu)$ is the microturbulent broadening kernel. Generally, this kernel is taken to be an isotropic zero-mean Gaussian with standard deviation $\xi$ in velocity units. This additional broadening propagates into the calculation of $t_\nu$ in Equation~\ref{eq:tau_alpha}, and from there to the intensity and flux. \par 

In contrast to microturbulence, macroturbulence is conceptualized as parcels of atmosphere larger than the photon mean free path that have some characteristic velocity distribution. Consequently, macroturbulence broadens intensity profiles rather than the absorption coefficients:

\begin{equation}
    I^{\rm mac}_\nu = I_\nu \circledast \Theta (\Delta\nu),
\end{equation}

\noindent where $\Theta (\Delta\nu)$ is the macroturbulence convolution kernel. In the simplest case, macroturbulence is modeled as an isotropic, zero-mean Gaussian with standard deviation $\zeta$ in velocity units. 
For isotropic macroturbulence, the convolution with $\Theta (\Delta\nu)$ can be moved outside of the integral (Equation~\ref{eq:flux}) when calculating the emergent flux, since by construction it has no $\mu$ dependency:

\begin{align} \label{eq:outside_integral}
    \mathcal{F}^{\rm mac}_\nu(\tau_\nu =0) &= \int_{0}^{1}  \Theta (\Delta\nu) \circledast \mu   I_\nu(\tau_\nu=0, \mu)d\mu \\ 
    \nonumber &= \Theta (\Delta\nu) \circledast \int_{0}^{1} \mu I_\nu(\tau_\nu=0, \mu)d\mu \\
    \nonumber &= \Theta (\Delta\nu) \circledast \mathcal{F}_\nu(\tau_\nu).
\end{align}

\noindent This analytical simplification expedites numerical evaluation of the flux, since it permits use of Equation~\ref{eq:milne}, obviating the numerical integration over $\mu$ (which necessitates evaluating the intensity at many positions on the disk). \par 




\subsubsection{Rotational Broadening} \label{subsubsec:rot_broadening}

Rotation broadens stellar spectra because local intensity profiles are Doppler shifted on the approaching and receding hemispheres of the star. This broadening is anisotropic by definition, since the line-of-sight component of the rotational velocity must vary with location on the disk due to projection, and if applicable, differential rotation. Mathematically, \citet{Gray2008} expresses the rotationally broadened flux as

\begin{equation} \label{eq:rot_integration}
    \mathcal{F}_\nu = \oint I_\nu (\Delta \nu - \Delta\nu
    _R) \cos \theta d\Omega,
\end{equation}

\noindent which denotes the integration of the local, rotationally shifted intensities over solid angle $\Omega$. Because the Doppler shifting of the line center ($\Delta \nu - \Delta\nu_R$) depends on location on the disk, the rotational broadening cannot be moved outside the integral as in Equation~\ref{eq:outside_integral}. Evaluating the above integral is somewhat computationally expensive, since it requires evaluating and integrating over many local intensities. However, \citet{Gray2008} show in their Equations (17.9)--(17.14) that, if it is assumed that center-to-limb variations in the continuum-normalized intensity profiles can be neglected (i.e., $I_\nu / I_{c}$ does not vary with location on the disk $\mu$), the continuum-normalized flux can be expressed as the convolution of the thermal profile $H(\Delta \nu) = I_\nu / I_c$ with a broadening kernel $G(\Delta \nu)$:

\begin{equation} \label{eq:rotation_convolution}
    \mathcal{F}_\nu / \mathcal{F}_c = H(\Delta \nu) \circledast G(\Delta \nu).
\end{equation}

\noindent The rotational broadening kernel $G(\Delta \nu)$ can be obtained by numerically integrating the line-of-sight velocities over the disk, or if linear limb darkening and solid-body rotation are assumed, it can be analytically derived (see \citealt{Gray2008} Equation (18.14)). An example rotational broadening kernel is shown as the solid orange curve in Figure~\ref{fig:kernels}. \par 

\subsubsection{Radial-Tangential Macroturbulence} \label{subsubsec:rt_broad}

In addition to rotation, which must be anisotropic, macroturbulence is also often modeled as an anisotropic broadening source. This is motivated by the fact that stellar convective cells (granules) are known to have some characteristic surface-horizontal and surface-vertical velocity, which will vary due to projection across the disk. Following from this picture, the canonical radial-tangential macroturbulence broadening kernel is given by \citet{Gray2008} as

\begin{align} \label{eq:full_rt_kernel}
    \Theta(\Delta \nu) =  & \frac{A_{\mathrm{R}}}{\pi^{1 / 2} \zeta_{\mathrm{R}} \cos \theta} e^{-\left(\Delta \nu/ \zeta_{\mathrm{R}} \cos \theta\right)^2} + \\ 
    \nonumber & \frac{A_{\mathrm{T}}}{\pi^{1 / 2} \zeta_{\mathrm{T}} \sin \theta} e^{-\left(\Delta \nu /\zeta_{\mathrm{T}} \sin \theta\right)^2},
\end{align}

\noindent where $\zeta_R$ and $\zeta_T$ are the characteristic radial and tangential broadening velocities, and $A_R$ and $A_T$ are the photon fractions from the radial and tangential flows, respectively. Generally, it is assumed that $\zeta_R = \zeta_T$ and $A_R = A_T$. \par 

In the above formulation, the $\theta$ dependence is explicit. However, \citet{Gray2008} show that (again under the assumption that center-to-limb variations in the intensity profiles can be neglected) this broadening can be expressed as a convolution:

\begin{equation} \label{eq:rt_macro_conv}
    \mathcal{F}_\nu = I_\nu \circledast M(\Delta \nu).
\end{equation}

\noindent The radial-tangential broadening kernel $M(\Delta \nu)$ is given by

\begin{align}
    \label{eq:iso_rt}
    M(\Delta \nu) = & \frac{2 A_{\mathrm{R}} \Delta \nu}{\pi^{1 / 2} \zeta_{\mathrm{R}}^2} \int_0^{\zeta_{\mathrm{R}} / \Delta \nu} \mathrm{e}^{-1 / u^2} \mathrm{~d} u\  + \\
    \nonumber & \frac{2 A_{\mathrm{T}} \Delta \nu}{\pi^{1 / 2} \zeta_{\mathrm{T}}^2} \int_0^{\zeta_{\mathrm{T}} / \Delta \nu} \mathrm{e}^{-1 / w^2} \mathrm{~d} w,
\end{align}

\noindent where $u = \zeta_R \cos\theta / \Delta\nu$ and $w = \zeta_T \sin\theta / \Delta\nu$. An example of this radial-tangential broadening kernel is shown as the dashed blue curve in Figure~\ref{fig:kernels}. We note that, in this formulation, 
the $\theta$ dependence has been analytically integrated out (see \citealt{Gray2008} Equations (17.7)--(17.8) and associated text). \par 

Since both rotational (Equation~\ref{eq:rotation_convolution}) and radial-tangential (Equation~\ref{eq:rt_macro_conv}) broadening can be expressed as convolutions under certain assumptions, \added{their broadening} can be modeled as a single convolution with a kernel that incorporates both broadening sources\footnote{\citet{Gray2008} show that it is incorrect to sequentially convolve the intensity with one kernel followed by the other; a monolithic kernel is required.}. \citet{Hirano2011} derive such a kernel for the case where $A_T = A_R$ and $\zeta_R = \zeta_T = \zeta_{\rm RT}$, assuming solid-body rotation and quadratic limb darkening (see their Equation~(B12), which gives the Fourier transform of the convolution kernel). An example rotation and radial-tangential broadening kernel (hereafter, ``the \citealt{Hirano2011} kernel'') is shown as the green dashed--dotted curve in Figure~\ref{fig:kernels}. \par 

\subsubsection{Accuracy of the Convolution Approximation} \label{subsubsec:conv_approx_error}

\begin{figure*}
    \epsscale{1.5}

    \gridline{\fig{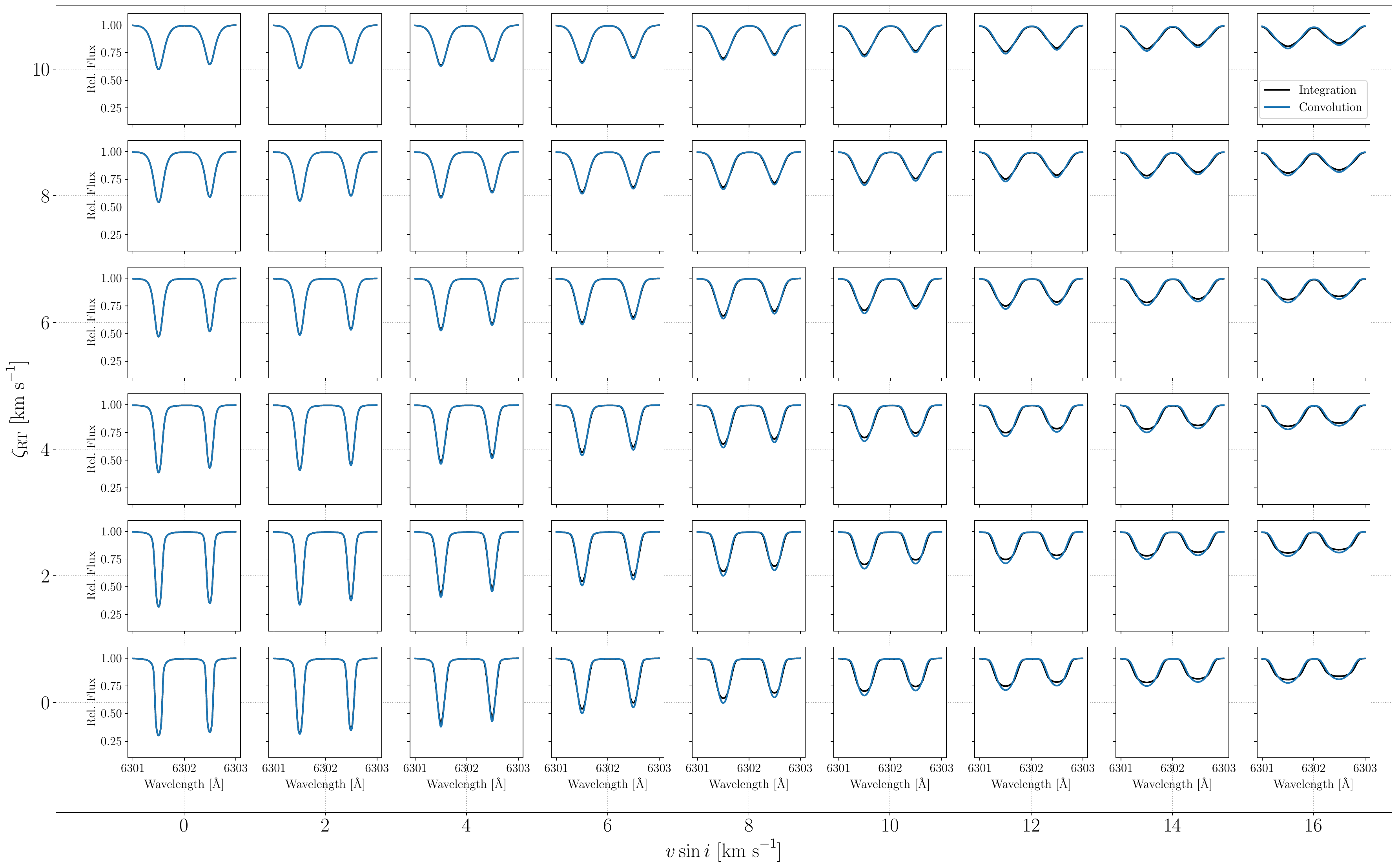}{0.925\textwidth}{}}
    \gridline{\fig{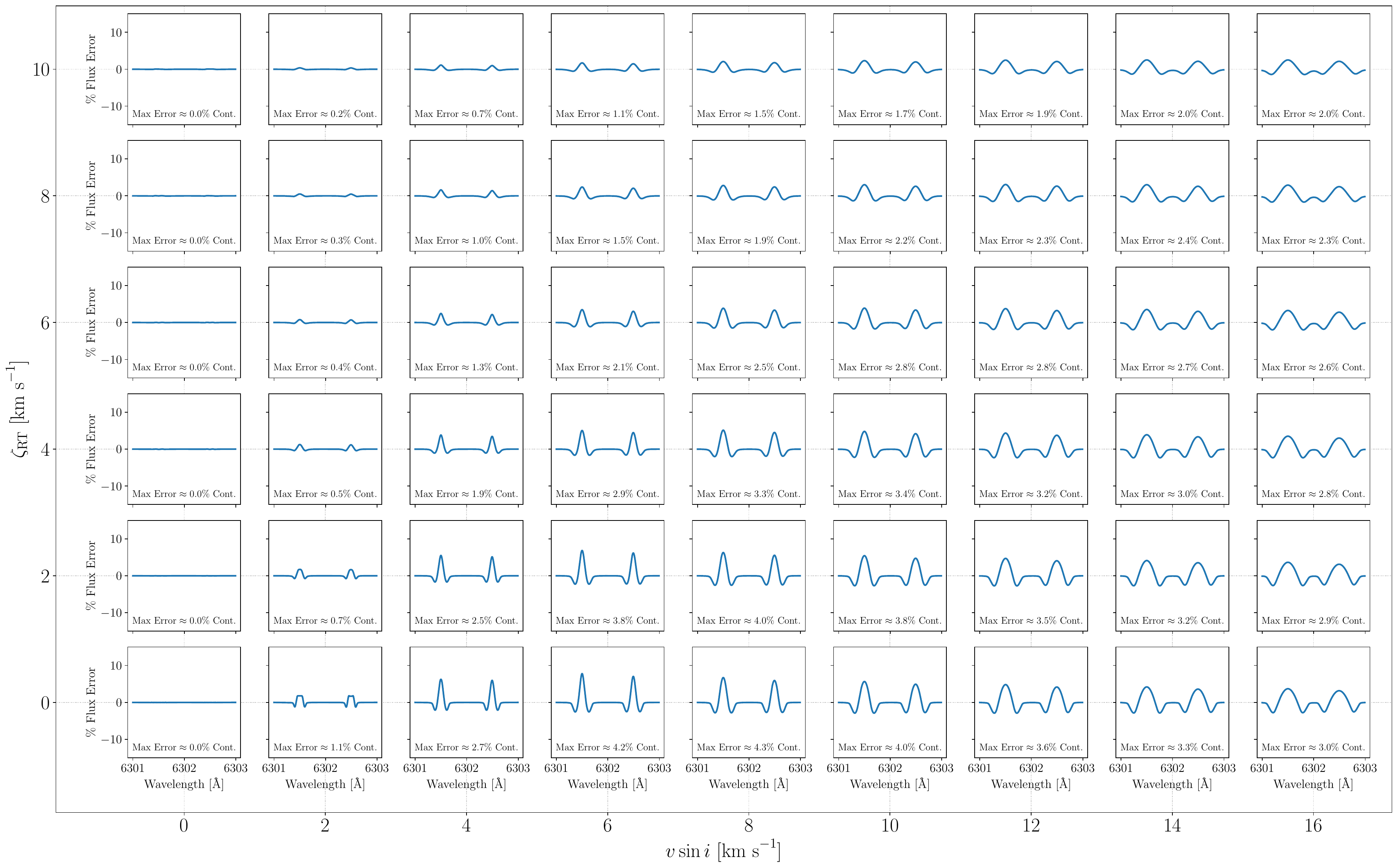}{0.925\textwidth}{}}

    \caption{The convolution method for broadening model stellar spectra can create appreciable errors in line shape, especially for stars with high $v \sin i$. \textit{Top:} flux computed via convolution (blue curves) and explicit disk integration (black curves) for various combinations of $v \sin i$ and $\zeta_{\rm RT}$. \textit{Bottom:} percent error in flux (expressed as a percentage of the disk-integrated continuum flux).}  \label{fig:big_plots_flux}
\end{figure*}

\begin{figure} 
    \epsscale{1.15}
    \plotone{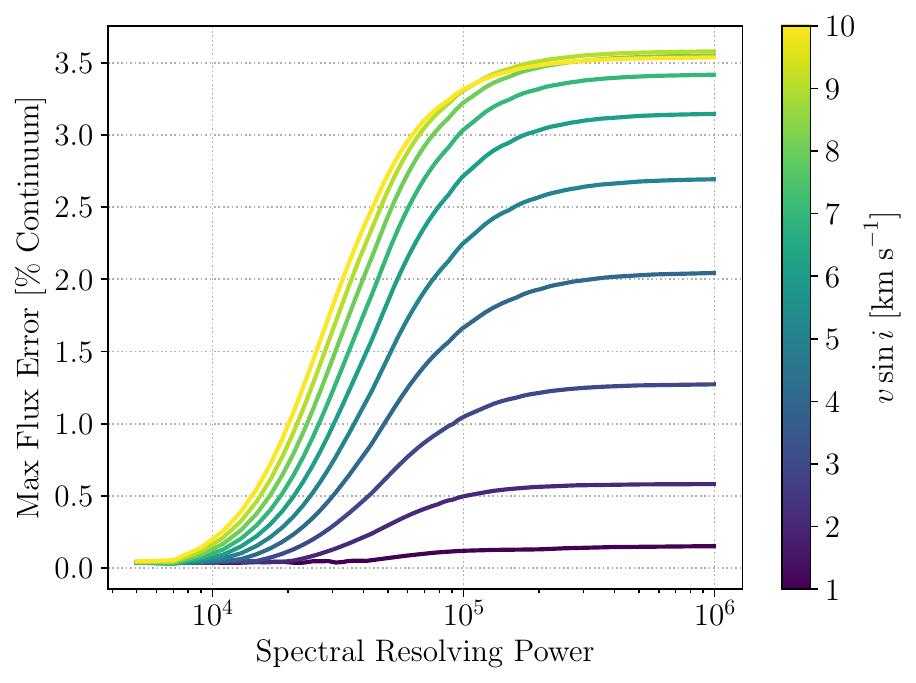}
    \caption{The flux error introduced by the convolution approximation decreases at lower spectral resolution. However, the error remains appreciable at the spectral resolving powers characteristic of modern EPRV instruments.} \label{fig:resolution}
\end{figure}

We have noted repeatedly in \S\ref{subsubsec:rot_broadening} and \S\ref{subsubsec:rt_broad} that the emergent flux spectrum can only be modeled via convolution under certain limiting assumptions that can break down in many science cases. First, the assumption that center-to-limb variations can be neglected is not generally valid and has, for example, led to erroneous detections of sodium in exoplanet atmospheres \citep[e.g.,][]{Casasayas-Barris2020}. Next, the \citet{Gray2008} and \citet{Hirano2011} broadening kernels assume solid-body rotation; stars are known to exhibit differential rotation, and this effect is detectable in analyses of stellar line shapes during planetary transits \citep[e.g.,][]{Doyle2023}\footnote{\added{The effect of differential rotation can be quite small, especially for fast rotators, and the need to include this effect can be science-case dependent (it is of notable importance in ``reloaded Rossiter--McLaughlin'' modeling; i.e., \citealt{Cegla2016}.)}}. Third, limb darkening is known to vary chromatically \added{across large bandpasses \citep{Neckel1994} and within line profiles, effects} that cannot be accounted for via convolution with a single broadening kernel. \par 

Because the above assumptions do not generally hold, the flux profiles modeled by 
convolution must incur some error that depends on the rotational velocity ($v\sin i$), the radial-tangential macroturbulent velocity broadening ($\zeta_{\rm RT}$), and more generally the intensity profile variations for a given model stellar atmosphere. The natural solution to this problem is to evaluate the flux by explicitly integrating Equation~\ref{eq:flux} over a model stellar disk, incorporating local variations in the rotational velocity and macroturbulent broadening in the local intensity profile. Though computationally more expensive, this approach allows one to include center-to-limb variations, non-solid-body rotation, and chromatic limb darkening dependence. To demonstrate differences in the convolution and integration formulations of the flux, we here compare their resulting flux spectra, varying the amount of rotational and macroturbulent broadening. Our procedures for producing these model spectra are described below. \par 

For both the convolved and integrated flux spectra, we use a MARCS \citep{Gustafsson2008} model solar atmosphere ($T_{\rm eff}$ = 5777 K, $\log g $ = 4.44) with microturbulent broadening of $\xi = 1.4\ \kms$. We adopt the solar abundances from \citet{Asplund2021}. To inspect line shape changes, we consider a $\sim$2 \AA\ region containing two Fe I lines. Atomic parameters for these lines are given in Table~\ref{tab:line_props}. Line and continuous absorption coefficients are calculated using \texttt{Korg.jl} \citep{Wheeler2023, Wheeler2024}. \added{In order to resolve all structure in the line profiles, we model the spectrum on a uniform wavelength grid with resolution 5$\times$10$^{-3}$ \AA\ and no instrumental broadening profile}.  \par 

To produce the convolved flux spectrum, we first generate the ``stationary'' spectrum from Equation~\ref{eq:milne} \added{(using $S_\nu(t_\nu) = B_\nu(T(t_\nu))$ with $t_\nu$ from Equation~\ref{eq:tau_alpha} and the temperature profile, $T$,  prescribed by the model atmosphere)}. This flux spectrum is then convolved with the \citet{Hirano2011} kernel, which accounts for the effects of both radial-tangential macroturbulence and rotation, assuming quadratic limb darkening and solid-body rotation. \added{To obtain the limb darkening parameters used in the kernel, we synthesized intensity profiles at many limb positions, and took the best-fit quadratic limb darkening coefficients for the continuum near the modeled lines ($u_1 \approx 0.465 $, $u_2 \approx 0.165$).} Since quadratic limb darkening laws notoriously fail near the limb, these parameters were determined for intensities evaluated at $\mu \geq 0.3$. \par 

To evaluate the disk-integrated flux, we first construct a model stellar surface grid following the tiling scheme developed by \citet{Vogt1987} and \citet{Piskunov2002}. At each tile on the grid, we evaluate the local intensity \added{(using Equations~\ref{eq:intensity} and \ref{eq:tau_alpha})} given the $\mu$ and rotational velocity at the center of the tile, which has the effect of simply Doppler shifting the line centers. Although it is possible to include differential rotation, we here implement solid-body rotation in order to compare as closely as possible to the convolution case, which assumes solid-body rotation. We then broaden each local intensity by the anisotropic form of the radial-tangential kernel given in Equation~\ref{eq:full_rt_kernel}, assuming that $A_R = A_T = 0.5$ and $\zeta_R = \zeta_T = \zeta_{\rm RT}$. Finally, we integrate these local intensities over the visible hemisphere of the star, yielding the emergent flux spectrum. \par 

We compare the flux computed via convolution to that computed by explicit disk-integration, as described above, for various combinations of $v\sin i$ and $\zeta_{\rm RT}$ in Figure~\ref{fig:big_plots_flux}. In addition to the flux spectra (top panel), we also show the relative errors in flux, taking the integrated flux as the fiducial (bottom panel). As expected, the error in line shape generally increases with either increasing $v \sin i$ or $\zeta_{\rm RT}$, although more strongly so with $v \sin i$. \added{At fixed $v \sin i$, the quoted \textit{maximum} flux error across the line profiles decreases with increasing $\zeta_{\rm RT}$. This is because the added broadening smooths the sharply peaked line core where the single largest flux discrepancy occurs. Redistributing this flux into the wings lowers the maximum error, but the cumulative error across the profile grows.}\par  


\subsection{Effects of Instrumental Spectral Resolution} \label{subsec:lsf}

\begin{figure*}[!ht]
    \epsscale{1.185}
    \plotone{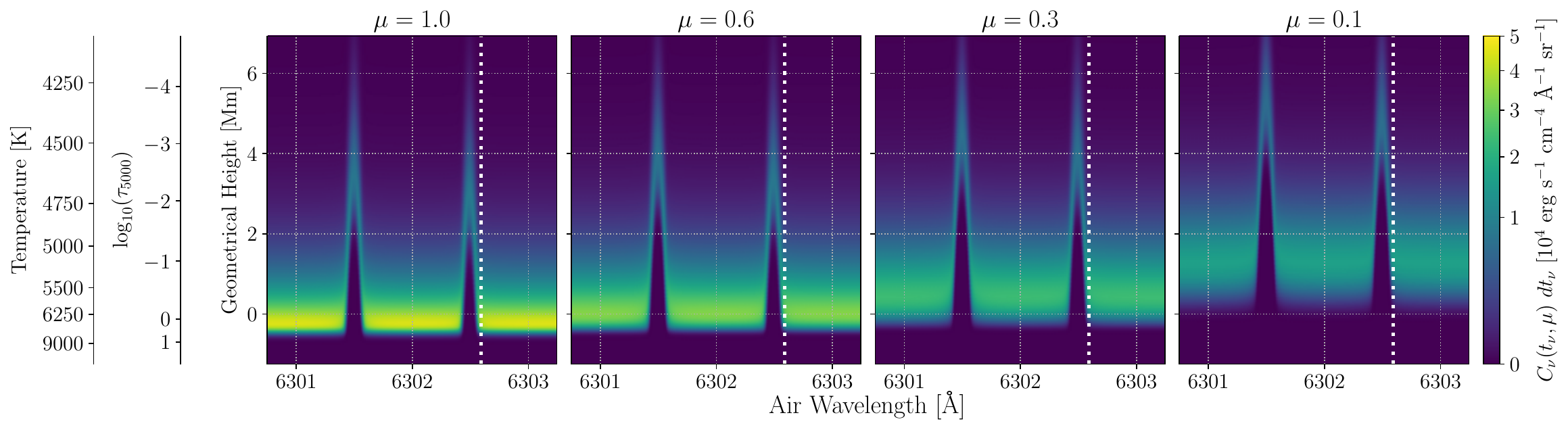}
    \caption{Example intensity contribution functions (multiplied by the differential $d t_\nu$) for four lines of sight ranging from disk center (leftmost panel) to near the limb (rightmost panel). Three coordinates from the model stellar atmosphere are shown for the vertical axis: geometrical height, optical depth at 5000 \AA, and temperature. At disk center, the continuum and lines form deeper in the atmosphere, compared to lines of sight nearer the limb. The white dotted vertical lines indicate the wavelengths at which slices through the intensity contribution functions are plotted in Figure~\ref{fig:cont_slice}.}
    \label{fig:cfuncs_intensity}
\end{figure*}

Since spectrographs have finite spectral resolving power $R$, with an associated line-spread function (LSF), and discretely sample the spectrum on a grid of pixels, the resolvable differences in flux obtained via convolution and integration should be diminished at lower spectral resolution. To quantify this expectation, we compare convolved and integrated model spectra consisting of the same solar Fe I lines, varying only $v \sin i$ and the spectral resolving power. We initially simulated the spectra with a constant wavelength resolution of $5 \times 10^{-3}$ \AA, before convolving with a Gaussian model LSF with standard deviation in velocity units given by the considered resolving power, $\Delta v = c / R$. We then performed a flux-conserving resampling of each spectrum onto a wavelength grid with 5 pixels per LSF FWHM \citep[following the algorithm presented in][]{Carnall2017}. \par 

The results of this exercise are shown in Figure~\ref{fig:resolution}. Following our expectation, the maximum flux error decreases with decreasing spectral resolving power, because the errors introduced by the convolution approximation are effectively ``blurred out'' by the LSF. At higher resolving powers ($R>10^4$), and especially for stars with higher $v \sin i$, the flux error remains appreciable ($\gtrsim$1 \%). This result indicates that spectral lines measured by modern EPRV spectrographs, which operate in the $R\gtrsim 10^5$ regime, will have mismodeled fluxes (and thus line shapes) at the few-percent level if they are computed with the convolution method. \par 

\section{Contribution Functions and Formation Temperatures} \label{contribution_functions}

In addition to the flux emergent from the top of the stellar atmosphere, it can also be useful in some contexts to consider which layers of a stellar atmosphere contribute to the shapes of absorption features. The integrands of Equation~\ref{eq:intensity} and Equation~\ref{eq:flux} provide such a mapping, and are consequently known as the intensity and flux ``contribution functions,'' respectively \citep{Rutten2003, Gray2008}. \par 

In many applications of stellar spectroscopic modeling, the contribution function is simply a mathematical intermediary needed to compute the intensity or flux spectrum. For other analyses, the contribution function is used more explicitly to determine the formation ``locations'' for absorption lines. \citet{AlMoulla2022}, for example, used contribution functions to determine the temperature in the model atmosphere of a star at which $50\%$ of the normalized, cumulative contribution function is reached. They dubbed this quantity $T_{1/2}$, and provided a visual representation of this definition in their Figure~1 (see also the left-hand plot of our Figure~\ref{fig:cum_cfunc_comp}). \par 


In their work, \citet{AlMoulla2022} used the intensity contribution function to model $T_{1/2}$. At a given $t_\nu$ coordinate in the stellar atmosphere, the intensity contribution function $C_\nu(t_\nu, \mu)$ is given by the integrand of Equation~\ref{eq:intensity}:

\begin{equation} \label{eq:int_cont}
    C_\nu(t_\nu, \mu) \equiv \frac{1}{\mu} S_\nu(t_\nu) \mathrm{e}^{-t_\nu/\mu}.
\end{equation}

\noindent The intensity contribution function $C_\nu(t_\nu, \mu)$ has the same dimensions as specific intensity -- energy per time per area per frequency per solid angle (e.g., erg s$^{-1}$ cm$^{-2}$ Hz$^{-1}$ sr$^{-1}$ in cgs). We can also define the cumulative intensity contribution function, which \citet{AlMoulla2022} used to define $T_{1/2}$, as the running integral of $C_\nu$ with respect to the optical depth $t_\nu$:

\begin{align}
    C_\nu^{\rm cum}(t_\nu, \mu) &= \int_0^{t_\nu} C_\nu(t_\nu, \mu) d t_\nu \\
    \nonumber &= \int_0^{t_\nu} \frac{1}{\mu} S_\nu(t_\nu) \mathrm{e}^{-t_\nu/\mu} d t_\nu.
\end{align}

Because the intensity contribution function is a function of $\mu$, it must vary with location on the disk. To demonstrate this effect, we plot example intensity contribution functions (multiplied by the differential $dt_\nu$) in Figure~\ref{fig:cfuncs_intensity} for multiple values of $\mu$. It is clearly visible in each panel that, for lines of sight closer to the limb (i.e., lower $\mu$ values), both the absorption features and continuum form higher in the atmosphere. \par 

As for the intensity, we can also define a contribution function for the flux. In its simplest form (we will discuss complications to the flux contribution function in \S\ref{subsec:contribution_complication}), the flux contribution function can be defined as the integrand of Equation~\ref{eq:milne}:

\begin{equation} \label{eq:flux_cont}
    \mathscr{C}_\nu(t_\nu) \equiv S_\nu(t_\nu) E_2(t_\nu),
\end{equation}

\noindent where we use a script $\mathscr{C}$ to mirror the convention of the script $\mathcal{F}$ used for flux. The flux contribution function $\mathscr{C}_\nu(t_\nu)$ has the same dimensions as specific flux -- energy per time per area per frequency (e.g., erg s$^{-1}$ cm$^{-2}$ Hz$^{-1}$ in cgs). The cumulative flux contribution function is then given by the following running integral: 

\begin{align}
    \mathscr{C}^{\rm cum}_\nu(t_\nu) = \int_0^{t_\nu} S_\nu(t_\nu) E_2(t_\nu) d t_\nu.
\end{align}

\par 

\begin{figure}
    \epsscale{1.15}
    \plotone{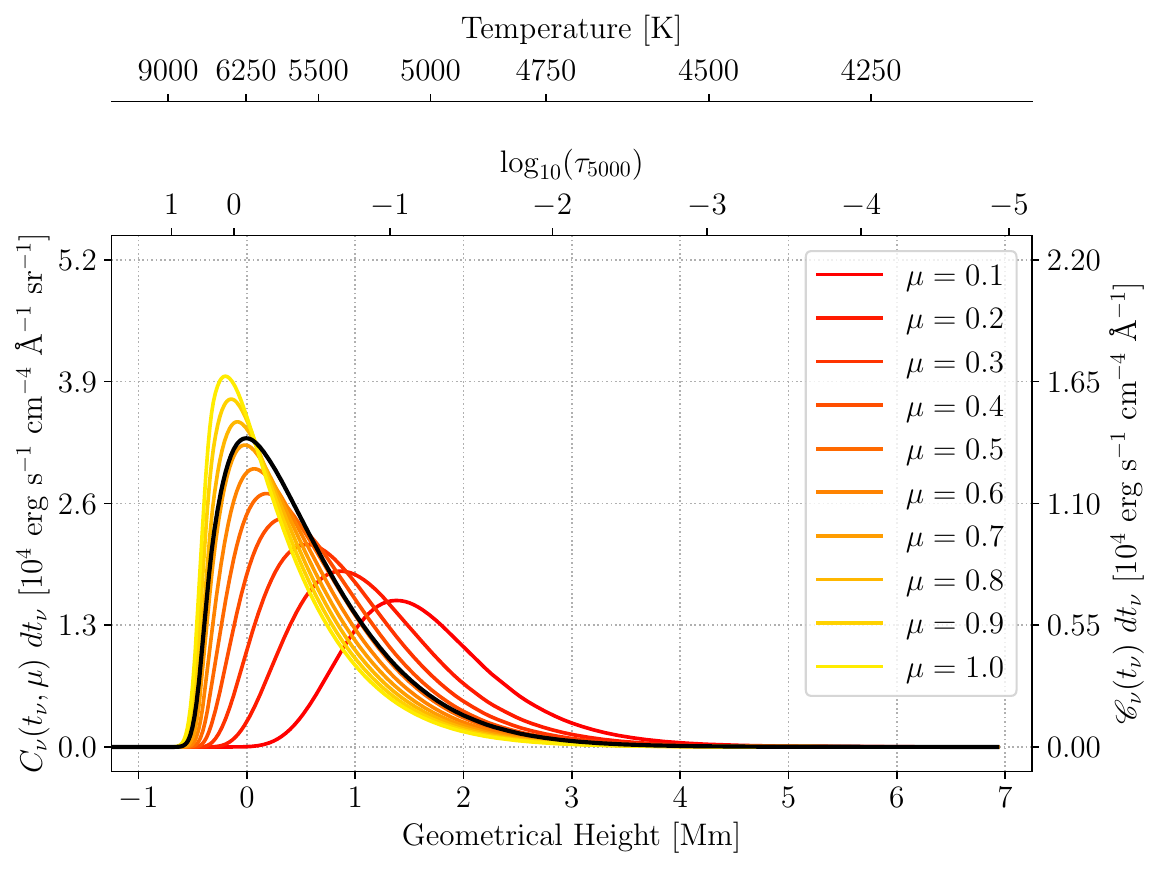}
    \caption{Example intensity (left axis, colored curves) and flux (right axis, black curve) contribution functions (multiplied by the differential $d t_\nu$) for a single wavelength element denoted by the vertical dotted lines in Figure~\ref{fig:cfuncs_intensity}. Three coordinates from the model stellar atmosphere are shown for the horizontal axis, like for Figure~\ref{fig:cfuncs_intensity}. Though intensity and flux have different dimensions (and consequently the scaling between the left and right axes is arbitrary), they are here overplotted to emphasize the differences in their shape over the height of model stellar atmosphere. Specifically, the intensity contribution shifts higher in the atmosphere with decreasing $\mu$, and the flux contribution covers a comparatively wide swath of atmosphere.}
    \label{fig:cont_slice}
\end{figure}

Just like flux integrates intensity over the disk (weighting by $\mu$), the flux contribution function accounts for the contribution along many lines of sight over the disk. This is shown in Figure~\ref{fig:cont_slice}. As was shown in Figure~\ref{fig:cfuncs_intensity}, the intensity contribution functions shift higher in the atmosphere as $\mu$ tends toward the limb. The flux contribution function, by contrast, is qualitatively more widely distributed over the full height of the atmosphere than any one intensity contribution function, as it effectively represents the (weighted) disk average. \par 

\subsection{Intensity vs.\ Flux and Formation Temperature} \label{subsubsec:int_vs_flux}

Because of the appreciable differences in shape between the various intensity contribution functions and the flux contribution function, it is important to use the correct expression when modeling formation temperatures or other similar quantities. Generally, when modeling other stars (or disk-integrated sunlight), it is correct to use the flux contribution function, since the stellar surface is not resolved. Notably, \citet{AlMoulla2022} use the \textit{intensity contribution function} (their Equation (2)) evaluated at $\mu = 1.0$ (i.e., disk center) to calculate formation temperatures, erroneously referring to it as the ``flux contribution function.'' For the HARPS-N solar spectra and HARPS $\alpha$ Cen B spectra they are modeling, their contribution function does not account for the contribution of higher (cooler) layers of the stellar atmospheres to the line profile. Consequently, the formation temperatures calculated from their expression are artificially hot. \par

To demonstrate this, we plot the cumulative intensity and flux contribution functions, as well as formation temperature spectra for the same model Fe I lines from Table~\ref{tab:line_props}, in Figure~\ref{fig:cum_cfunc_comp}. In the continuum, the $\mu = 1.0$ formation temperature spectrum is visibly offset toward higher temperatures compared to that calculated from the cumulative flux contribution function. In the line cores, the differences between the intensity and flux $T_{1/2}$ values is more complicated. To highlight this, we show the ``errors'' in $T_{1/2}$ (the difference between the $\mu = 1$ intensity $T_{1/2}$ values and the flux $T_{1/2}$ values) in the bottom panel of the right-hand plot in Figure~\ref{fig:cum_cfunc_comp}. Clearly, the differences in formation temperature driven by the \added{substitution of the flux contribution function for the intensity contribution function} are not a simple offset, but also create line-by-line shape differences in the formation temperature spectrum. \par 




\subsection{Effects of Additional Nonthermal Broadening on the Contribution Functions and Formation Temperatures} \label{subsec:contribution_complication}

\begin{figure*}
    \epsscale{1.15}
    \plottwo{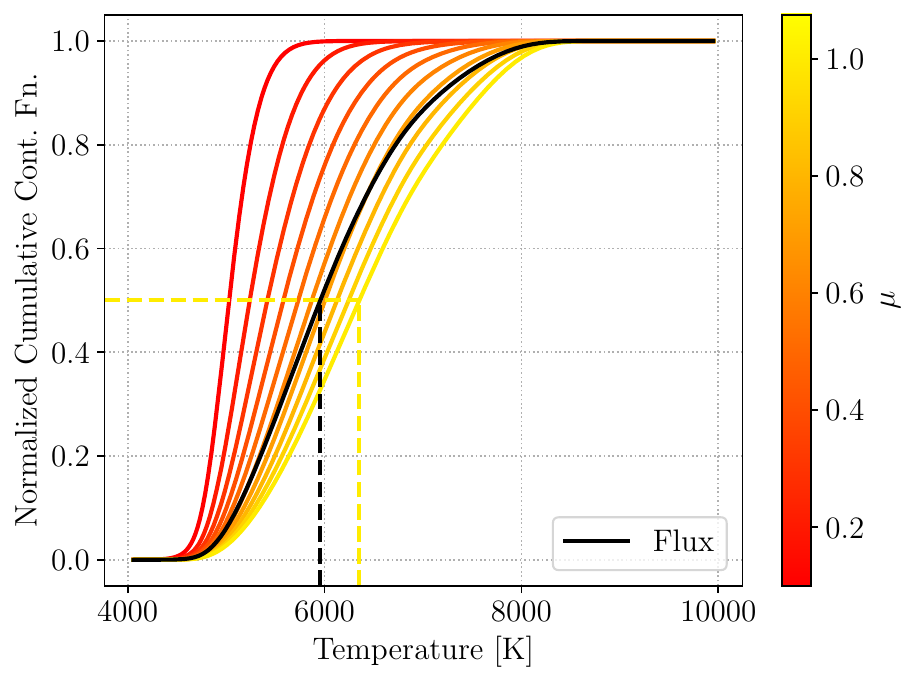}{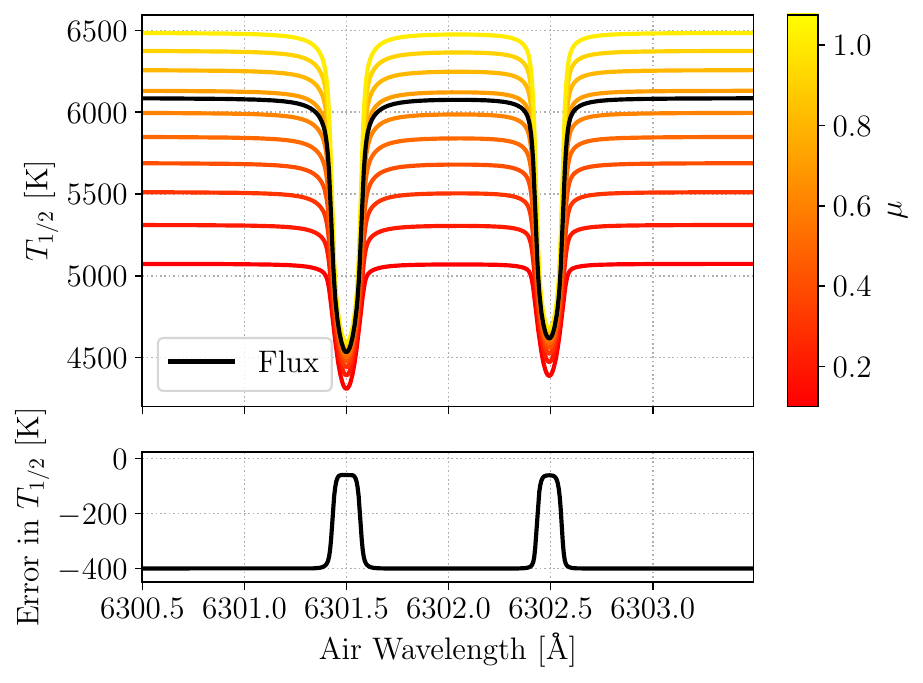}
    \caption{Normalized, cumulative contribution functions and formation temperatures are very sensitive to the distinction between intensity and flux. \textit{Left:} normalized, cumulative intensity (colored curves) and flux (black curve) contribution functions. The vertical and horizontal dashed lines indicate the formation temperature $T_{1/2}$ for flux (black) and disk-center intensity (yellow). \textit{Right:} formation temperature spectra for intensity at many values of $\mu$ (colored curves) and flux (black curve). The flux formation temperature is markedly cooler than the intensity formation temperature for $\mu = 1$. The difference between the formation temperature calculated from the disk-center intensity contribution function and the flux contribution function is shown in the bottom panel. The difference is not a constant offset; there is a marked difference in shape in the line cores.}
    \label{fig:cum_cfunc_comp}
\end{figure*}

Because sources of additional broadening (e.g., microturbulence, macroturbulence, and rotation) affect the intensity and flux, they must also affect the intensity and flux contribution functions. Their impact on the intensity contribution is simple to consider. As previously discussed in \S\ref{subsubsec:isotropic}, the effects of microturbulence are ``built into'' the intensity contribution function, since isotropic microturbulence broadens the line opacities/absorption coefficients directly. \added{Rotation is accounted for in intensity profiles by Doppler shifting line centers given the local, line-of-sight projected rotational velocity.} Macroturbulence, as treated in both \S\ref{subsubsec:isotropic} and \S\ref{subsubsec:rt_broad}, broadens local intensities directly. It is mathematically identical to perform this convolution within the intensity integral:

\begin{align}
        I^{\rm mac}_\nu &= I_\nu \circledast \Theta (\Delta\nu) \\
        \nonumber &= \int_0^{\infty} \Theta (\Delta\nu) \circledast  \frac{1}{\mu} S_\nu\left(t_\nu\right) \mathrm{e}^{-t_\nu / \mu} \mathrm{d} t_\nu \\
        \nonumber &= \int_0^{\infty} \Theta (\Delta\nu) \circledast  C_\nu(t_\nu, \mu) \mathrm{d} t_\nu.
\end{align}

\noindent The final integrand above gives the macroturbulently broadened intensity contribution function. In situations where one is only interested in the emergent intensity, this is computationally inefficient, since it requires performing a convolution for each layer in the stellar atmosphere model, rather than one convolution after evaluating the emergent intensity integral. In cases where the full contribution function is of interest (e.g., in calculating formation temperatures), the above formulation is appropriate, although slower to evaluate. \par 

Appropriately treating the effects of \added{additional} nonthermal broadening on the flux contribution function likewise necessitates some excess computation beyond what is simply required for obtaining the broadened flux. Just like the flux is the weighted integral of intensity over the disk, the flux contribution function can be expressed as the weighted integral of the intensity contribution functions over the disk. This broadened flux contribution function can either be computed via the convolution approximation or by explicit disk-integration. \par  

\begin{figure*} 
    \gridline{\fig{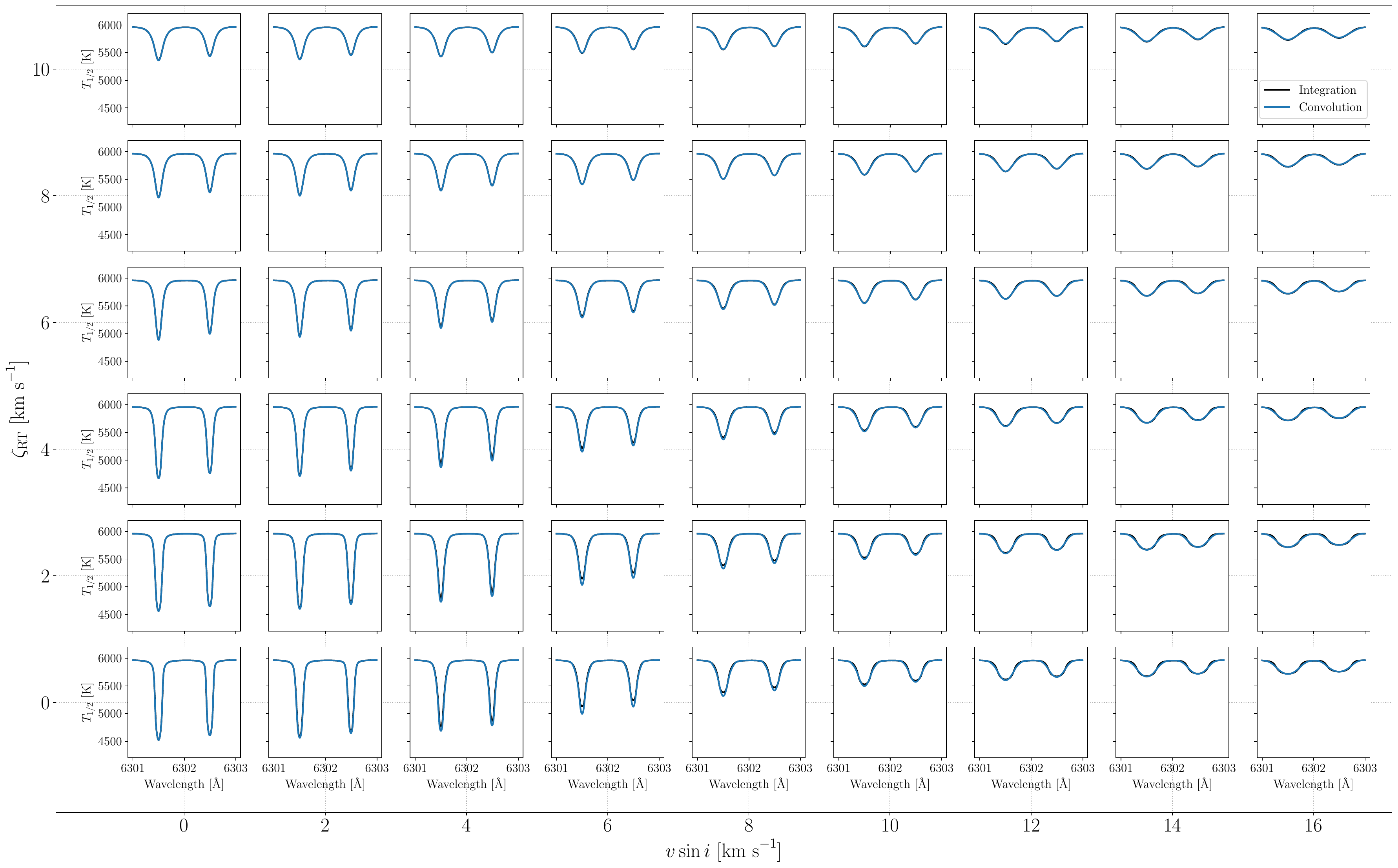}{0.925\textwidth}{}}
    \gridline{\fig{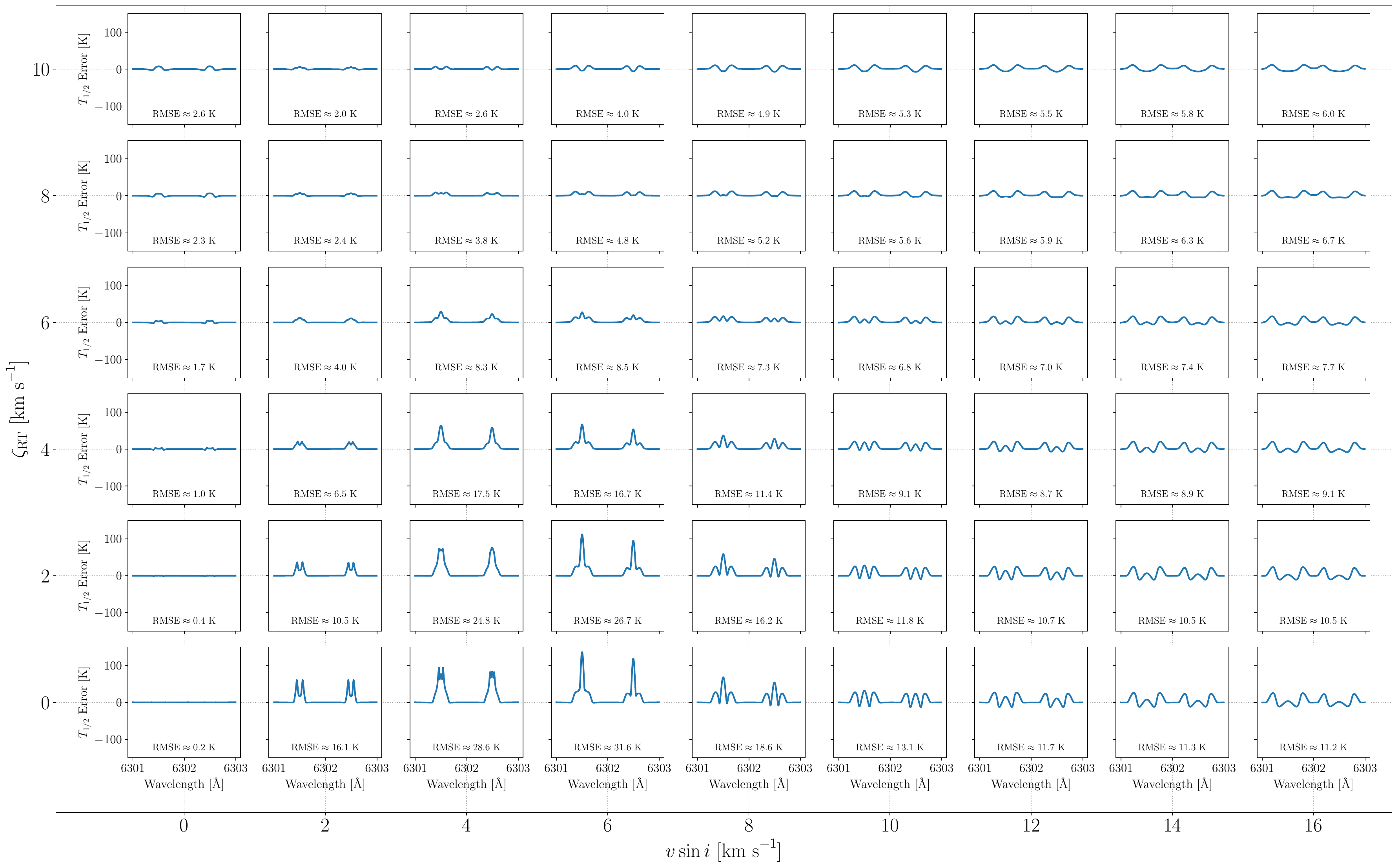}{0.925\textwidth}{}}
    
    \caption{Similar to Figure~\ref{fig:big_plots_flux}, but for formation temperature $T_{1/2}$ instead of flux. \textit{Top:} formation temperature obtained via convolution (blue curves) and disk integration (black curves) for various combinations of $v \sin i$ and $\zeta_{\rm RT}$. \textit{Bottom:} error in $T_{1/2}$.} \label{fig:big_plots_temp}
\end{figure*}

To quantify how the convolution approximation applied to the flux contribution function affects the modeled formation temperatures, we repeat the exercise performed in \S\ref{subsubsec:conv_approx_error} and presented in Figure~\ref{fig:big_plots_flux}, instead for the formation temperature. These results are presented in Figure~\ref{fig:big_plots_temp}. In the top panel, we show the formation temperature spectra computed via convolution and integration for varying $v \sin i $ and $\zeta_{\rm RT}$. In the bottom panel, we show the error in formation temperature. Like for the flux errors shown in Figure~\ref{fig:big_plots_flux}, the degree of error in formation temperature is primarily driven by $v \sin i$. For a solar-like star ($v \sin i \sim 2.1 \kms$ and $\zeta_{\rm RT} \sim 3 \kms$), the root mean square error (RMSE) in $T_{1/2}$ is only about several Kelvin. At worst, the error is a few tens of Kelvin, small compared to the error produced by \added{using} the intensity contribution function \added{instead of the flux contribution function} (upwards of hundreds of Kelvin in the continuum, and dozens in line cores; see Figure~\ref{fig:cum_cfunc_comp}). \par 

\section{Discussion and Implications} \label{discussion}

We have shown in the previous sections that modeling spectral line broadening and computing formation temperatures are nontrivially impacted by modeling choices. Here, we further discuss the potential impacts of these findings and consider in which science cases they may (or may not) be relevant. We also caveat limitations of the formation temperature formalism and the assumptions present in our analyses. \par 

\subsection{Flux Errors across a Stellar Sample}

\begin{figure} 
    \epsscale{1.15}
    \plotone{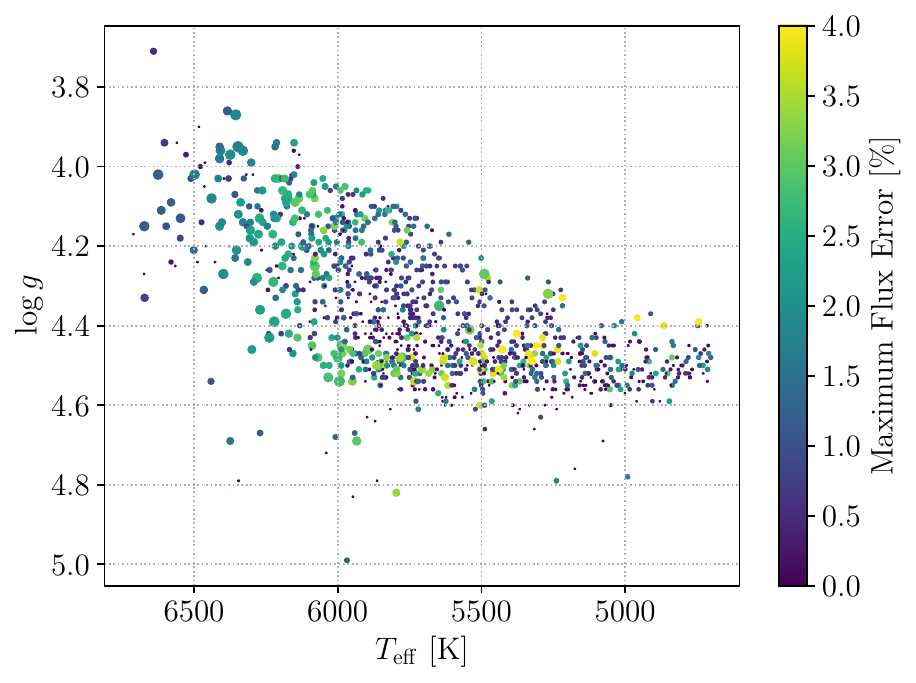}
    \caption{\added{Flux errors incurred by modeling broadening via convolutions can vary strongly across parameter space. The color of points is given by the maximum difference between the flux computed by convolution and integration for a sample of stars with parameters measured in \citet{Brewer2016, Brewer2017}.} The size of the points are scaled to the reported $v \sin i$. As seen in Figure~\ref{fig:big_plots_flux}, rotation is the primary driver of flux error.
    } \label{fig:hr}
\end{figure}

In \S\ref{subsec:additional_broadening} and \S\ref{subsec:lsf}, we considered the effects of $\zeta_{\rm RT}$, $v \sin i$, and spectral resolving power $R$ on modeled line shapes. These, however, are only three of a myriad of properties that control the shapes of lines in stellar spectrum. To illustrate how the full gamut of stellar properties can interact with the convolution modeling assumption, we consider the stellar sample from \citet{Brewer2016, Brewer2017}, who homogeneously measure $T_{\rm eff}$, $\log g$, $v \sin i$, $v_{\rm mac}$, and $[{\rm Fe/H}]$ for a large sample of stars. Following \citet{Brewer2018}, we exclude stars that are likely to have evolved off the main sequence. \par  

For each star, we synthesize the convolved and integrated model spectra consisting of the same two Fe I lines from Table~\ref{tab:line_props}. \added{We synthesize on the same high-resolution (5$\times$10$^{-3}$\AA) grid as used for Figures~\ref{fig:big_plots_flux} and \ref{fig:big_plots_temp}.} We use stellar atmosphere models interpolated from the MARCS grid \citep[][]{Gustafsson2008}, with \added{elemental abundances computed from the reported $[\mathrm{Fe/H}]$ values (see \citealt{Wheeler2023} and \citealt{Wheeler2024} for details on abundance tabulations given metallicities)}. For the microturbulent broadening velocity $\xi$, we adopt the following parameterization from \citet{Bruntt2010}:

\begin{align}
    \xi(T_{\rm eff}) =\ & 2.75 \times 10^{-7} (T_{\rm eff} - 5777)^2\ + \\
    \notag & 4.56 \times 10^{-4} (T_{\rm eff} - 5777)\ +\ 1.01,
\end{align}

\noindent where $T_{\rm eff}$ is in Kelvin and $\xi$ in $\kms$. \par 

The errors between the convolved and integrated flux models are shown in Figure~\ref{fig:hr}. \added{Generally, the points with larger errors have larger values of $v\sin i$ (shown by the size of the scatter point).} Relatedly, the \added{density} of stars with appreciable flux errors increases above $\gtrsim$6000 K, corresponding roughly to the Kraft break where magnetic braking becomes inefficient \citep[][]{Kraft1965} \added{and consequently stellar rotational velocities increase.} \par 


\added{In generating Figure~\ref{fig:hr}, we have taken the stellar parameters reported by \citet{Brewer2016, Brewer2017} and the microturbulence from the \citet{Bruntt2010} relation as fiducial. \citet{Brewer2016} and \citet{Brewer2017} use Spectroscopy Made Easy (SME; \citealt{Valenti1996}) to derive stellar parameters; since SME uses a quadrature-based disk-integration scheme to produce stellar flux spectra, the stellar parameters reported by these works should not be impacted by modeling errors related to broadening. However, there is a large natural scatter in the microturbulence relation reported by \citet{Bruntt2010}, as seen in their Figure~11; we do not account for this scatter in this simple exercise. Consequently, we do not claim that the flux errors shown in Figure~\ref{fig:hr} are accurate for individual stars in the \citet{Brewer2016, Brewer2017} sample.} Rather, we have performed this exercise to show (broadly speaking) how the numerous stellar properties that influence the observed spectrum (e.g., abundances, rotational velocities, effective temperatures, etc.) will affect the modeling error incurred by using the convolution approximation. A detailed analysis of how the model flux errors in turn bias inferred stellar properties is beyond the scope of this work, but is one important next point of investigation. \par 

\subsection{Pitfalls of the Formation Temperature Model}

\begin{figure*}
    \gridline{\fig{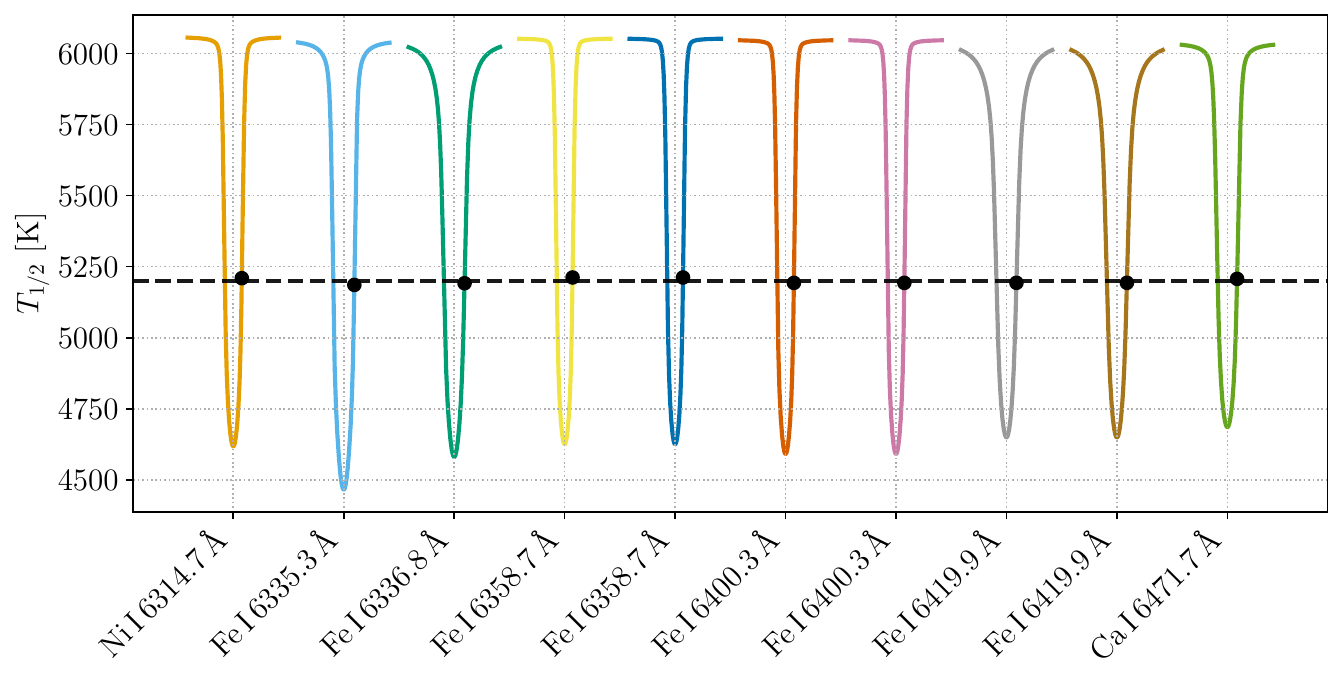}{0.45\textwidth}{(a)} 
              \fig{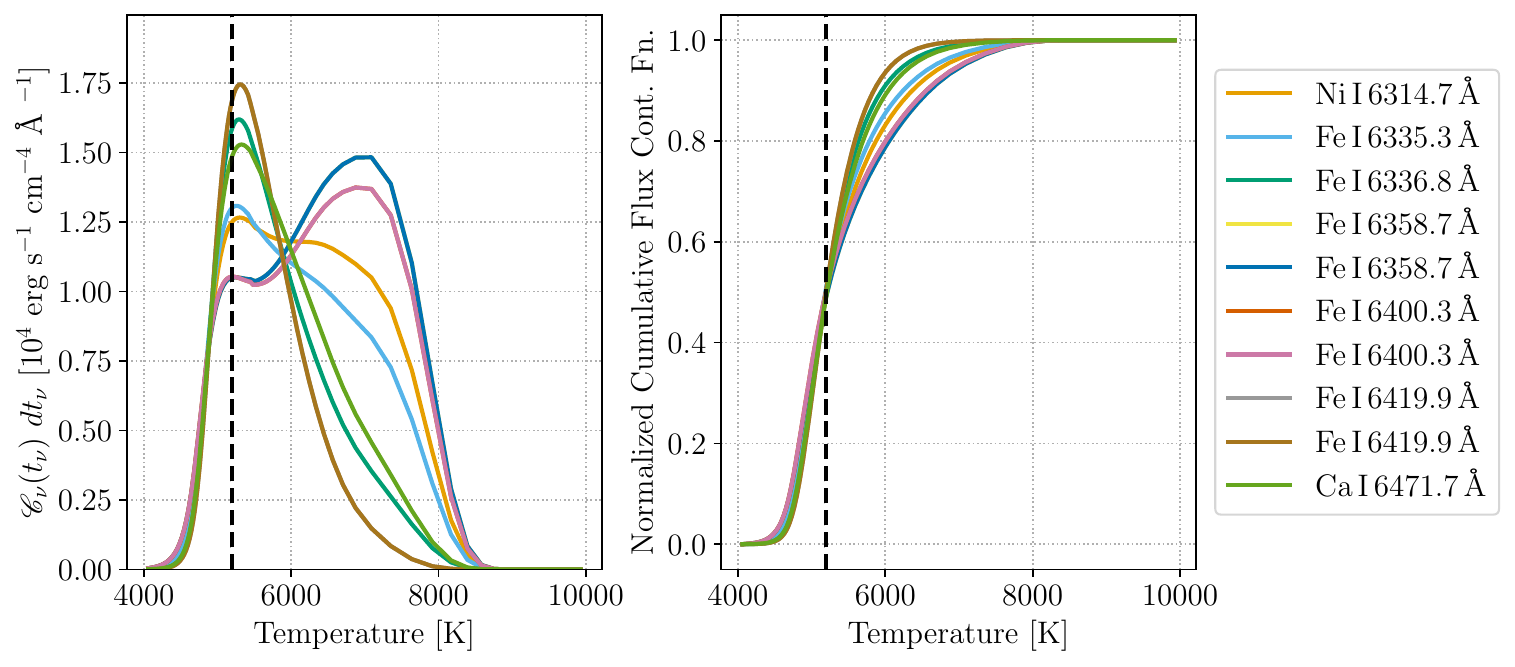}{0.525\textwidth}{(b)}}
    \caption{Spectral elements with the same formation temperature do not necessarily trace the same physical portions of the stellar atmosphere. \textit{Panel (a):} several arbitrarily chosen lines with at least one wavelength element in their red wing with a formation temperature of about 5200 K (denoted by the black points). \textit{Panel (b):} the contribution functions (multiplied by the differential $dt_\nu$, left) and normalized cumulative contribution functions (right) for each pixel denoted by the black points in the left-hand plot. Although these spectral elements all share a formation temperature, their contribution functions differ dramatically.}
    \label{fig:mean_distribution}
\end{figure*}

In \S\ref{subsubsec:int_vs_flux}, we showed that conflating the intensity contribution function with the flux contribution dramatically changes the modeled formation temperatures. We emphasize that flux, intensity, and their contribution functions are mathematically and physically distinct quantities, despite some tendency in colloquial language (and some places in the literature) to treat the terms ``flux'' and ``intensity'' as synonymous or otherwise interchangeable. Even properly accounting for this distinction, it bears emphasizing that formation temperatures are a simplification of the physics of radiative transfer. Formally, the very concept of assigning a singular formation temperature to \added{a piece} of a spectrum is incorrect; the contribution functions in Figure~\ref{fig:cont_slice} clearly show that light, in general, emerges from an extended region of the stellar atmosphere. Perhaps more insidiously, formation temperature spectra can lead to the false impression that \added{pieces of the spectrum} with the same formation temperature trace the same ``regions'' of the stellar atmosphere; their contribution functions may cover markedly different distributions over height (and local gas velocities) but just so happen to share a similar 50$^{\rm th}$ percentile. \par 

To illustrate the dangers of this assumption, we synthesized several line profiles and computed formation temperature spectra for various chemical species in a solar model atmosphere. These line profiles (in formation temperature) are shown at left in Figure~\ref{fig:mean_distribution}. For each of these line profiles, we also plot the flux contribution functions for the wavelength elements whose formation temperature is closest to $\sim$5325 K at right in Figure~\ref{fig:mean_distribution}. Although these portions of the modeled spectrum have essentially the same formation temperature, their contribution functions are quite different in certain cases. Future works may consider how line classifications based on other summary statistics of the contribution function (or indeed the full contribution functions themselves) affect the inferred intrinsic variability in EPRV spectra. \par 

\subsection{Limiting Assumptions in This Analysis} \label{sec:assumptions}

Throughout this work, we have made a number of assumptions that are quite common in stellar radiative transfer. Namely, we have assumed that our atmosphere is homogeneous, plane-parallel, and in LTE. Of course, these assumptions are false in real stellar atmospheres. For dwarfs, which we are principally concerned with, the stellar atmosphere is very thin compared to the radius of the star, such that the plane-parallel assumption is reasonable. This will not be true of \added{stars where the stellar radius is not very large compared to the pressure scale height (e.g., highly evolved stars)}, and so we caution that the results presented in this work (and the software accompanying) should not be applied in \added{such} cases. \par 

The assumption of LTE is also prone to failure. It is widely known that non-LTE physics can significantly sculpt absorption lines, especially those that form in the chromosphere \citep[e.g.,][]{Lagae2025}. As in \citet{AlMoulla2022}, we recommend rejecting lines that are clearly poorly modeled by the synthesis and those that are known \textit{a priori} to be sensitive to NLTE effects. \par 

In \S\ref{subsec:additional_broadening}, we noted that the derivation of Equation~\ref{eq:rotation_convolution} required neglecting center-to-limb variability. In the most general sense, the phrase \textit{center-to-limb variability}, or CLV, can refer to any of a myriad of effects that cause intensity spectra to vary with limb angle. This terminology is quite overloaded and can include (depending on field of study, author intent, etc.) effects such as limb darkening, line asymmetry changes, line depth changes, and variable convective blueshift (i.e., the ``limb shift''). Although the plane-parallel synthesis used herein yields simple line profile variations from center to limb (visible in Figures~\ref{fig:cfuncs_intensity} and \ref{fig:cum_cfunc_comp}), this model cannot produce variations that result from 3D convective motions in the stellar atmosphere. For that, state-of-the-art 3D (magneto)hydrodynamical simulations coupled with radiative transfer codes are needed, which have recently begun to reproduce with remarkable fidelity these precise variations in asymmetry \citep[e.g.,][]{Frame2025} for solar absorption lines. \par 

The fidelity achieved by \citet{Frame2025} is possible because their simulations reproduce the vertical and horizontal velocity gradients created by convection at the solar surface. These convective motions encode themselves in the shapes of absorption lines observed by high-resolution solar spectrographs. Future works, rather than attempting to simplistically model these motions (either with the 1D approach presented in this work, or the 1.5D approach employed by \citealt{Frame2025}) could instead attempt to infer these motions (and the resulting perturbed and shifted contribution functions) directly from observations of lines. These inferred velocities and the resulting contribution functions would be closer to measured quantities than the purely modeled ones presented in this work. \par 


\section{Conclusions} \label{conclusions}

In this work, we explore the consequences of modeling assumptions in accounting for the effects of nonthermal broadening (i.e., microturbulence, macroturbulence, and rotation) in stellar spectra. We show three key results:

\begin{enumerate}
    \item Accounting for macroturbulence and rotation via a convolution can lead to appreciable flux errors, especially in the case of high $v\sin i$. 
    \item Modeled formation temperatures are \textit{very} sensitive to the distinction between intensity and flux. On top of this distinction, proper treatment of rotation and macroturbulence further perturbs the modeled formation temperatures.
    \item Formation temperatures are a form of summary statistic of the contribution function. Consequently, formation temperatures can elide some complexities of radiative transfer if taken at face value. 
\end{enumerate}

The code (\texttt{FormationTemperatures.jl}) used to generate the figures and other quantitative results used in this work is available from GitHub\footnote{\url{https://github.com/palumbom/FormationTemps.jl}} and archived on Zenodo as \citet{FormationTemps_jl}. \par 

\begin{acknowledgments}
M.L.P.\ would like to warmly thank K.\ Al Moulla, M.\ Bedell, R.\ A.\ Rubenzahl, A.\ J.\ Wheeler, J.\ T.\ Wright, and the anonymous referee for insightful conversations and comments that have contributed to this manuscript. 
M.L.P.\ also thanks J.\ M.\ Brewer for his aid in screening likely evolved stars from the stellar sample shown in Figure~\ref{fig:hr}.
M.L.P.\ was supported by the Flatiron Research Fellowship at the Flatiron Institute, a division of the Simons Foundation. 
This work has made use of the VALD database, operated at Uppsala University, the Institute of Astronomy RAS in Moscow, and the University of Vienna. 
This research has made use of NASA's Astrophysics Data System Bibliographic Services. 
\end{acknowledgments}

\software{Julia \citep{Julia},
          \texttt{Korg.jl} \citep[][]{Wheeler2023, Wheeler2024},
          FormationTemps.jl \citep[][]{FormationTemps_jl},
          Matplotlib (\citealt{Hunter2007}),
          JuliaGPU \citep{JuliaGPU}}

\bibliography{main, misc}{}
\bibliographystyle{aasjournalv7}

\end{document}

%% file: spectroscopic.tex
\centerwidetable
\begin{deluxetable*}{ccccccc}
\tabletypesize{\footnotesize}
\tablecaption{Parameters for spectroscopic lines modeled in this work. \label{tab:line_props}}

\tablehead{\colhead{Species} & \colhead{Air Wavelength} & \colhead{$E_{\rm lower}$} & \colhead{$\log(gf)$} & \colhead{$\log \gamma_{\rm stark}$} & \colhead{$\log \gamma_{\rm rad}$} & \colhead{$\log \gamma_{\rm vdW}$} \\
& (\AA) & (eV) & & & & }

\startdata
Fe I & 6301.4995 & 3.6537 & -0.72 & -5.41 & 8.08 & -7.54 \\
Fe I & 6302.4931 & 3.6864 & -0.97 & -5.39 & 8.08 & -7.54 \\
\enddata
\tablecomments{Taken from the Vienna Atomic Line Database (VALD; \citealt{Piskunov1995}).}
\end{deluxetable*}